\documentstyle[aaspp4,epsfig]{article}
\def\paren#1{\left( #1 \right)}

\def\bra#1{\left[ #1 \right]}

\begin{document}

\title{Optical Flashes and Radio Flares in GRBs afterglow: Numerical Study}
\author{Shiho Kobayashi$^1$ and Re'em Sari$^2$}

\begin{abstract}
The strong optical flash observed by ROTSE, as well as the radio
flare associated with GRB 990123 can be attributed to the emission of the
fireball ejecta, initially heated by the reverse shock. 
We numerically study the evolution of an adiabatic relativistic
fireball interacting with an ambient uniform medium, both in the
initial energy transfer stage and in its late evolution. It is shown
that the Blandford-McKee solution adequately describes the evolution
of the shocked shell quite early on and for as long as the fireball
material has relativistic temperatures. In the case where the reverse
shock is only mildly relativistic the shocked shell becomes cold
almost immediately and the evolution deviates from the Blandford-McKee
solution. We derive analytical expressions for the ejecta evolution in
its cold regime. This solution gives a good approximation to the
numerical results.
We estimate the radiation from the fireball ejecta using the
numerical hydrodynamic evolution in both cases: cold and hot
shells. Surprisingly, we find that both evolutions give rather similar
light curves, decaying approximately as $T^{-2}$ in the optical and peaking
at about one day in the radio, even though the hydrodynamics is different.
\end{abstract}

\affil{$^{1}$Department of Earth and Space Science, Osaka University,
Toyonaka, Osaka 560, Japan\newline
$^{2}$Theoretical Astrophysics 130-33, California Institute of Technology,
Pasadena, CA 91125, USA}




\section{Introduction}

Precise and prompt gamma-ray burst localization by an Italian-Dutch
satellite, BeppoSAX makes it possible for other space and ground
instruments to scan the same direction. The long-lasting counterparts
(afterglow) in X-ray, optical and radio wavelength have been detected. Up to
the event of GRB~990123, the overall behavior of the afterglows could be
explained reasonably well by the synchrotron emission from the ambient medium
particles (ISM) that are shocked by the relativistic flow.

GRB~990123, is the brightest gamma-ray burst seen
by BeppoSAX and the fluence is in the top $0.3\%$ of all bursts observed so
far (Kippen et al. 1999). Absorption lines in the optical afterglow gave a
lower limit of the redshift $z > 1.6$, the energy required to produce the
bright gamma-ray burst is enormous, $3\times10^{54}$ erg for an isotropic
emission. A new clue to understand the nature of gamma-ray bursts was found
in this event by ROTSE which detected a strong optical flash during the
``gamma-ray'' burst. ROTSE started to observe the direction in an optical
band 22 seconds after the onset of the burst. The optical flash reached to a
peak of 9th magnitude and then decayed with a slope of a power law index $%
\sim 2$ (Akerlof et al. 1999).

Such a strong prompt optical flash was predicted (M\'esz\'aros and Rees
1997, Sari and Piran 1999a,b). The prompt optical flash observed by ROTSE
is compatible with these predictions (Sari and Piran 1999c, M\'es\'aros and
Rees 1999). In the energy transfer stage of a fireball
evolution, the forward shocked ISM and the reverse shocked fireball shell
carry comparable amount of internal energy. However, the typical temperature
in the shocked ejecta is considerably lower than that of the shocked ISM.
Consequently, the typical frequency of the synchrotron emission from the
shocked shell is much lower than that from the shocked ISM, and it comes to
the optical band with reasonable values of the parameters.

Beside the optical flash, GRB~990123 had another newly observed
phenomenon, a radio flare (Kulkarni et al. 1999). While the usual
radio afterglows rise on time scale of weeks or months, this burst had
a radio flare peaking at one day, rising quickly before the peak and
decaying quickly after the peak. Sari and Piran (1999c) have
interpreted this flare as the emission from the ejecta particles 
earlier shocked by the reveres shock. The same particles producing 
the prompt optical flash have been cooled adiabatically and
their emission shifts quickly towards lower frequencies while
weakening. According to the analytical estimates of Sari and Piran,
scaling the prompt optical emission to the epoch of the radio
detection gives the right amount of radio emission.

The discussion of the prompt optical flush is quite robust, the
estimate of the hydrodynamical evolution of the shocked ejecta is
more fragile.  In this paper we study numerically the energy
transfer stage and the evolution of the ejecta after that in order to
estimate the decay rate of the optical flash as well as the light curve
and timing of the radio flare.  We consider both cases, where the
temperature of the shocked shell is relativistic or not. Using
equipartition assumption on the magnetic field we construct the light
curve of the emission from the shell.
In section 2, we shortly review the hydrodynamics of a fireball.
We show in section 3 the analytic estimate of the evolution of 
the shocked ejecta for both case of the hot and the cold shell case.
The radiation from the ejecta
are also estimated. In section 4, we discuss the numerical
results. Finally, in section 5 we summarize the results and
discuss their implications.


\section{Hydrodynamics of a Fireball}


The evolution of an adiabatic spherical fireball with an energy $E$ and a
dimensionless entropy (or random Lorentz factor) $\eta$ and a radius $R_0$
is characterized by several phases (Kobayashi, Piran and Sari, 1999).
Initially, the fireball expands into a surrounding medium with a low density 
$\rho_1$, a narrow shell with a radial width of $\Delta \sim R_0$ is formed.
The Lorentz factor of the shell increases linearly with the radius during
the free acceleration stage. At $R_L \equiv R_0 \eta$ the fireball shell
uses up all its initial radiation energy, it coasts with the Lorentz factor
of $\eta$ and the frozen radial width $\Delta \sim R_0$. 
The coasting ends
once the ISM begins to influence the shell. 
The interaction between the
shell and the ISM is described by two shocks: a forward shock
propagating into the ISM and a reverse shock propagating into the
shell.

A dimensionless quantity $\xi \equiv (l/\Delta)^{1/2}/\eta^{4/3}$ is useful
to consider the evolution of the fireball shell after the coasting stage
(Sari and Piran 1995) where $l \equiv (E/\rho_1c^2)^{1/3}$ is the Sedov length.
The evolution can be classified into two categories by this quantity. 
We call the case of $\xi >1$
the Newtonian Reverse Shock (NRS) case, and $\xi < 1$ the Relativistic
Reverse Shock (RRS) case.

If initially $\xi >1$ (the NRS case), the coasting shell begins to
spread as $\Delta \sim R/\eta^2$ at $R_s \equiv R_0\eta^2$ due to a
slight difference of the velocity inside the shell. 
The reverse shock is initially Newtonian and becomes mildly
relativistic when the reverse shock crosses the shell.  After the
crossing which happens at $R_\gamma \equiv l/\eta^{2/3}$, the forward
shocked ISM has the most energy of the system.

If initially $\xi <1$ (the RRS case), the reverse shock becomes
relativistic at $R_N=l^{3/2}/\Delta^{1/2}\eta^2$ which is before the
shock crosses the shell. At $R > R_N$ the reverse shock begins to
reduce considerably the Lorentz factor of the shell's matter which it
crosses. The distance between the contact discontinuity and the
reverse shock is $\propto R^{2}$ where $R$ is the forward shock radius. After the drastic deceleration,
during the shock crossing, the shocked shell slows down as 
$\propto R^{-1/2}$ due to the pressure difference between the
forward shock and the reverse shock.  
At $R_\Delta \equiv l^{3/4}\Delta^{1/4}$, the shock crosses
the whole shell and most of the energy of the system has been
transferred to the forward shocked ISM. The spreading effect is not
important in this case since the spreading radius $R_s$ is larger than
$R_\Delta$.

Kobayashi, Piran and Sari (1999) have shown numerically that after
$R_{\gamma}$ for the NRS case or $R_\Delta$ for the RRS case, the
profile of the shocked ISM begins to approach the Blandford-McKee
solution (Blandford and McKee 1976; BM hereafter) which is the basis
of most of the afterglow theories. The fraction of the energy of
the ejecta rapidly decays as the forward shocked ISM has most of
the energy in the system. We are interested in the evolution of this
reverse shocked ejecta as the source of the rapidly decaying optical
flash and the radio flare (Sari and Piran 1999c, Kulkarni et al. 1999).


\section{Late Ejecta Evolution - Analytical Estimates}

\subsection{Hydrodynamics - Relativistic Temperatures}


In the previous section we discussed the hydrodynamical evolution of
the ejecta up to the time that a considerable fraction of the energy
was given to the surrounding medium. Beyond this time, one has to
calculate the evolution of the Lorentz factor, the pressure (or the
energy density) and the mass density as functions of time in order to
evaluate the emission from the ejecta.  Sari and Piran (1999b,c) used
the Blandford-McKee solution to estimate all these quantities. Using
this solution, the Lorentz factor, the pressure and the mass density
of a fluid element are given by
\begin{equation}
\gamma \propto R^{-7/2}, p \propto R^{-26/3}, \rho \propto R^{-13/2}.
\label{bm_scalings}
\end{equation}
Since the observer time is given by $T \propto R/\gamma^2c \propto R^8$, we
obtain $\gamma \propto T^{-7/16}, p \propto T^{-13/12}$ and 
$\rho \propto T^{-13/16}$.

Sari and Piran (1999c) found good agreement between the light curves
derived with the above scalings and the observed optical flash and
radio flare. However, it is not clear whether the BM solution is
applicable to the reverse shocked ejecta for the following reasons. 
(i) The BM solution, as a self similar solution, describes the
shocked ISM long after the energy transfer stage where the details of
the initial conditions are no longer important. (ii) It assumes that
the initial shell is irrelevant, however the evolution of that shell
is what we are interested in. (iii) The initial shell contains much
more particles than those collected by the forward shock. Its density 
therefore must be higher than that predicted by the BM solution. 
(iv) Though the BM solution assumes relativistic temperatures, a
mildly relativistic reverse shock can not heat ejecta to such a high
temperature. 

It can be argued that since the shocked ejecta is located not too far
behind the forward shock at the end of the energy transfer stage and
it has a comparable amount of energy to that of the system, it roughly
fits the BM solution. The fact that its rest mass density is much higher
than that given by the BM solution should not play an important role
as long as the temperature is relativistic. At relativistic
temperature, the fluid inertia is due to its thermal energy rather
than the rest mass. The fluid therefore can be expected to evolve
according to the BM scalings with density that is higher by a
constant factor from that given by the BM solution. We will show
numerically that indeed the points (i)-(iii) raised above are not a
real problem, and the BM solution adequately describes the evolution
of the ejecta from the very early stage as long its temperature is
relativistic. The forth point however, i.e., the underlying assumption
of the BM solution that the temperature is relativistic might become
incorrect quite early on.  Specifically, if the reverse shock is only
mildly relativistic, this assumption breaks from the beginning and the
BM solution can not describe any of the ejecta evolution.

M\'{e}szaros and Rees (1999) assumed a general power law evolution of
the forward shock and approximated the ejecta Lorentz factor to be
equal to that of the forward shock, $\gamma \propto R^{-g}$. The
density was estimated by assuming that the ejecta spreads in the local
frame by the speed of light, the width in the local frame is therefore
$R/\gamma$. Since the number of particles there is constant, the
density drops as $n\propto R^{-3}\gamma \propto R^{-3-g}$.  They
considered two possible evolutions of the pressure. (I) The ``pressure
equilibrium'' assumption that the ejecta pressure is equal to the
forward shock pressure, leads to $p\propto \gamma ^{2}\propto
R^{-2g}$.  (II) Adiabatic expansion $p \propto
\rho^{4/3}$.  In terms of observed time these scalings are: $\gamma
\propto T^{-g/(1+2g)}$, $\rho \propto T^{-(3+g)/(1+2g)}$ and $p\propto
T^{-2g/(1+2g)}$ for the ``pressure equilibrium''and $p\propto
T^{-4(3+g)/3(1+2g)}$ for the ``adiabatic expansion''.

Though for a larger value of g the Lorentz factor decreases faster 
with observed time, it is relatively insensitive to $g$. Even
for $g\rightarrow \infty $ the Lorentz factor decreases as $T^{-1/2}$ 
while for the lowest reasonable value of $g=3/2$ it decays as
$T^{-3/8}$. For $g=7/2$ one obtains the same deceleration law of 
the Lorentz factor and the density as a fluid element in the BM
solution follows, i.e. the BM solution satisfies the spreading
assumption. Furthermore, the BM solution satisfies the adiabatic
expansion law, $p\propto \rho ^{4/3}$, but not the pressure equilibrium.
Note also
that though the power law $\gamma \propto R^{-g}$ agrees with the BM evolution
for a fluid element, it is not proportional to the Lorentz factor of the 
shock, which evolves as $R^{-3/2}$.

\subsection{Hydrodynamics - Sub-relativistic Temperatures}

Non of the two papers above considered the sub-relativistic
temperature regime. As mentioned earlier this regime can be important
quite early on if the reverse shock is mildly relativistic. It is
impossible to repeat the analysis of Sari and Piran (1999 a,b) since 
there is no known analytical solution describing this regime. The BM
solution is not applicable since the temperature is non relativistic
and the Sedov-Taylor solution is not applicable as the bulk Lorentz
factor of the fluid is relativistic. However, we can minimize the
uncertainty to a single parameter in a restricted range.

Assume as above that $\gamma \propto R^{-g}$. If $\gamma$ is not
described by a power law of $R$, a limited rage of radius is considered 
over which the value of $g$ is approximately a constant. We can expect
$g$ to be higher than $3/2$ since the ejecta must lag behind the
forward shock, but we can expect $g<7/2$ as compared to the BM
solution the ejecta has higher inertia (due to its non negligible rest
mass) and therefore it is expected to be slowed down less abruptly.

We can now use a version of the spreading assumption as follows: when
the ejecta arrives at radius $R$ the time in the local frame is
$R/c\gamma $.  The ejecta sound speed is sub-relativistic and can be
estimated by $(p/\rho)^{1/2}$. The width of the shell will therefore
be $(p/\rho)^{1/2}R/c\gamma$. The ejecta density is therefore: $\rho
\propto R^{-3}\gamma (\rho /p)^{1/2}$.  Using the adiabatic expansion
law $p\propto \rho ^{4/3}$, we get
\begin{equation}
\gamma \propto R^{-g}, p\propto R^{-8(3+g)/7}, 
\rho \propto R^{-6(3+g)/7}.
\label{subrela}
\end{equation}
In terms of the observer time these scalings become:
$\gamma \propto T^{-g/(1+2g)}$, $p\propto T^{-8(3+g)/7(1+2g)}$ and 
$\rho \propto T^{-6(3+g)/7(1+2g)}$. 

We note that the above derivation is not rigorous. A power law
evolution of $\gamma$ as function of radius is not necessarily correct
in the sub-relativistic temperature regime. Moreover, the spreading
assumption may not be valid if the pressure gradients in the ejecta
are steep. We therefore use the above only as a guiding line to
compare with the numerical results.

We assumed above (and through the paper) that the fluid is described
by a constant adiabatic index $\hat{ \gamma}=4/3$. This is clearly
true as long as the protons are relativistic. Once the shell is cold
and the protons are no longer relativistic, their adiabatic index
becomes $\hat{\gamma}=5/3$. A mixture of newtonian protons and
relativistic electron results in adiabatic index of $\hat \gamma =
13/9$ if the two species are kept in equipartition. However, the basic
assumption here is that some level of equipartition is only created by
the passage of a shock since the collision time is too long. Once the
proton become newtonian and their adiabatic index becomes $\hat
\gamma=5/3$, electrons and protons will deviate from equipartition
since they evolve differently. They always have the same density due
to the charge neutrality but the thermal energy of protons evolves as
$\rho^{5/3}$ while that of the electrons evolves as $\rho^{4/3}$. As
$\rho$ decreases, the relativistic electrons dominate the thermal
energy and the pressure of the fluid. Since the electrons are still
relativistic far after the protons become cold we assume that the
fluid is described by a constant adiabatic index $\hat \gamma =4/3$,
even after the shell become cold with newtonian protons.

\subsection{The Ejecta Emission}

The reverse shock propagates into the shell and heats its
electrons.  After it has crossed the shell, no new electrons are
injected.  The emission from the reverse shocked shell reaches the peak at
$R_\gamma$ for a NRS case ($\xi > 1$) or $R_\Delta$ for a RRS case
($\xi < 1$). Since the Lorentz factor of the shocked region at that
time is $\eta$ and $\xi^{3/4}\eta$ respectively, the peak time for the
observer is given by $\xi^2R_0/c$ and $R_0/c$ respectively. According 
to the internal shock model the duration of the gamma-ray burst itself
is $R_0/c$. 

After the peak time when the reverse shock crossed the shell, the
shocked electrons cool radiatively and adiabatically. We consider here
the simplest case in which the energy of the magnetic field 
remains a constant fraction of the internal energy $B^2 \propto p$. The
electron random Lorentz factor evolves as $\gamma _m \propto p/\rho $ due to
the adiabatic expansion. The typical synchrotron frequency in the observer
frame is $\nu _m \propto \gamma \gamma _m^{2}B$, the spectral power
at the typical frequency is $F_{\nu _{m}}\propto \gamma B$ for a fixed 
total number of radiating electrons. Assuming a power law distribution 
of the electron random Lorentz factor with index $\hat p$, the spectral
flux at a given frequency above $\nu _{m}$ is $F_{\nu }\sim F_{\nu
_{m}}(\nu /\nu _{m})^{-(\hat p-1)/2}$ while below $\nu_m$ we have the
synchrotron low energy tail as 
$F_{\nu }\sim F_{\nu_{m}}(\nu /\nu _{m})^{1/3}$.
Substituting the expressions for $\nu_m$ and $F_{\nu_m}$ we have 
\begin{equation}
\label{emission} 
F_{\nu} \propto \left\{
                \begin{array}{@{\,}ll}
    \gamma ^{2/3}p^{-1/3}\rho^{2/3}            & \nu < \nu_m  \\          
    \gamma ^{(\hat p+1)/2}p^{(5\hat p-3)/4}\rho^{-(\hat p-1)} & \nu > \nu_m.
                \end{array}
                \right. 
\label{flux}
\end{equation}
This is a generalized form of equation 3 in Sari and Piran (1999c) in which
they substituted the BM relations and $\hat{p}=2.5$ to get a $T^{-2.1}$
decay above $\nu_m$ and a $T^{-17/36}$ below $\nu_m$. 

If the power law scalings (\ref{subrela}) are valid, typical 
frequency $\nu_m$ and the peak flux $F_{\nu_m}$ evolve as $\nu_m \propto
T^{-3(8+5g)/7(1+2g)}$ and $F_{\nu_m} \propto T^{-(12+11g)/7(1+2g)}$. The
flux at a frequency below $\nu_m$ (above $\nu_m$) drops as
$T^{-2(2+3g)/7(1+2g)}$ ($T^{-(7+24\hat{p}+15\hat{p}g)/14(1+2g)}$). These
decay indexes of the flux are monotonic functions of $g$ and are not 
so sensitive to it. If the value of $g$ is limited as
$3/2< g < 7/2$, these vary in relatively narrow rages for $\hat
p=2.5$. The index for the low frequency part is between $-0.46$ 
and $-0.44$, the index for the high  part is between $-2.2$ and $-1.8$.
These are very close to the estimates from the BM scalings.

At low frequencies and early times, self absorption takes an important
role and significantly reduces the flux. A simple estimate of the
maximal flux is the emission from the black body with the reverse
shock temperature. The temperature is given by the random energy of 
the typical electron $m_e c^2\gamma_m$ for a frequency below the typical
frequency $\nu_m$. If the observed radio frequency is above it, 
the electron radiating into the observed frequency has
energy higher by a factor $(\nu/\nu_m)^{1/2}$ since the synchrotron 
emission frequency is proportional to the square of the Lorentz factor.
Using the same expression as in Sari and Piran (1999c), but leaving 
arbitrary the hydrodynamic evolution we get an upper limit to the 
emission of 
\begin{equation}
F_{\nu} \cong 5.0 \times 10^{-8} {\rm [Jy]} \gamma^3 \gamma_e
\max\bra{1,\paren{\frac{\nu}{\nu_m}}^{1/2}}\paren{\frac{T}{\mbox{1day}}}^{2}
\paren{\frac{\nu}{\mbox{8.5 GHz}}}^2.
\label{bb}
\end{equation}
where we assumed $\Omega_0=1, \lambda_0=0$, $h=0.65$ and the location
of the fireball $z=1.6$ as GRB990123.
The emission will therefore be the minimum between that
given by equation (\ref{emission}) and equation (\ref{bb}).

Since our numerical simulation is purely hydrodynamic, we can not use
them to verify any of the radiation assumptions leading to equations
(\ref{emission}) and (\ref{bb}). We will therefore use the above expression
to evaluate the output radiation from the hydrodynamical properties of 
the ejecta. However, we will be able to get a more realistic
hydrodynamic by using the numerical simulations.

\section{Numerical Simulation}

The initial configuration for our simulation is a static uniform
spherical fireball surrounded by a uniform cold ISM. It is determined
by four parameters: the total energy $E$, the dimensionless entropy
$\eta$, the initial radius $R_0$ and the ISM density $\rho_1$. $E$ and
$\rho_1$ always appear as the ratio of $E/\rho_1$ in the hydrodynamics
computation, the system is actually determined by three parameters,
the initial radius $R_0$, the entropy $\eta$ and the Sedov length $l$.
First we consider two extreme cases: of the RRS $\xi << 1$ and the 
NRS $\xi >>1$ to see the difference clearly. Then, the case of
GRB~990123 will be studied.


\subsection{the Relativistic Reverse Shock Case}

The temperature of the reverse shocked ejecta is one of the 
differences between the RRS case and the NRS case. We suspect
that the relativistic temperature is a crucial condition to 
apply the BM solution to the shocked ejecta. We consider 
a relativistic reverse shock case $E=3\times10^{54}$erg, 
$\rho_1=10$ proton $\mbox{cm}^{-3}$, $\eta=2\times10^5$ and 
$R_0=3\times 10^{10}$cm, and we compare the evolution of the
ejecta with the BM solution. This parameter set corresponds  
to $\xi=10^{-3}$. 

For computational efficiency, this simulation is
started at $R_\Delta/100 \sim 5\times 10^{14}$ cm at which the ejecta
shell is in the coasting stage. It is larger than $R_N \sim 2 \times
10^{12}$ cm where the reverse shock becomes relativistic, but at that 
time only $10^{-6}$ of the ISM material within $R_\Delta$ had been
swept up. In other word, the reverse shock had decelerated $10^{-4}$ 
of the shell (Sari and Piran 1995; Kobayashi, Piran and Sari 1999). 
Therefore, the deceleration prior to this time can be neglect.

\begin{figure}[b!] 
\centerline{\epsfig{file=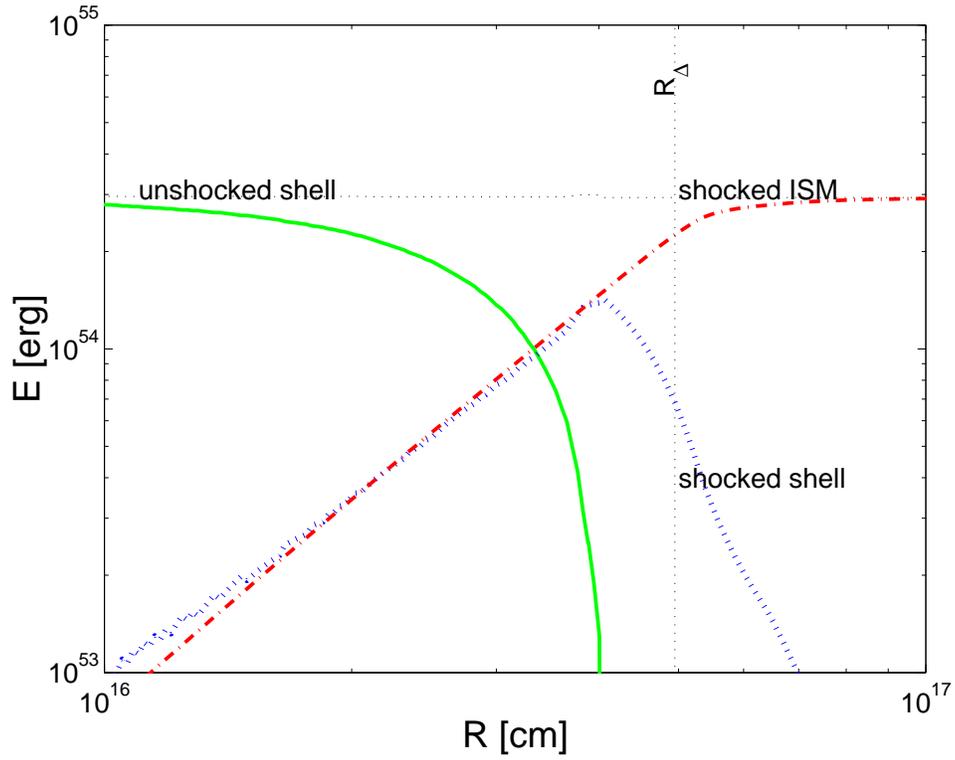,width=5in}} \vspace{10pt}
\caption{Energy transfer from the fireball to the shocked regions. The 
sum of the kinetic and the thermal energies in the unshocked shell
(solid line), in the reverse shocked shell (dotted line)
and in the forward shocked ISM (dashed dotted line). The horizontal
dotted line is the sum of the energies in the three region.
$R_\Delta$ (vertical dotted line) is a analytic estimate where the reverse
shock crosses the ejecta.} 
\label{fig:energy_transfer}
\end{figure}

Initially, the unshocked fireball shell has all the energy of the
system. As the shell expands, the reverse shock decelerates the ejecta
while the forward shock accelerate the ISM. The energy is transferred from 
the unshocked shell to the ISM via the shocks, finally the shocked ISM
carries all the energy of the system. In the intermediate stage, around
$R_\Delta$, the shocked shell has comparable energy to the shocked ISM.
The evolutions of the energies in three regions, inside of the reverse
shock (unshocked shell), between the reverse shock and the contact
discontinuity (shocked shell) and between the contact discontinuity and
the forward shock (shocked ISM) are shown in figure
\ref{fig:energy_transfer}. 

\begin{figure}[b!] 
\centerline{\epsfig{file=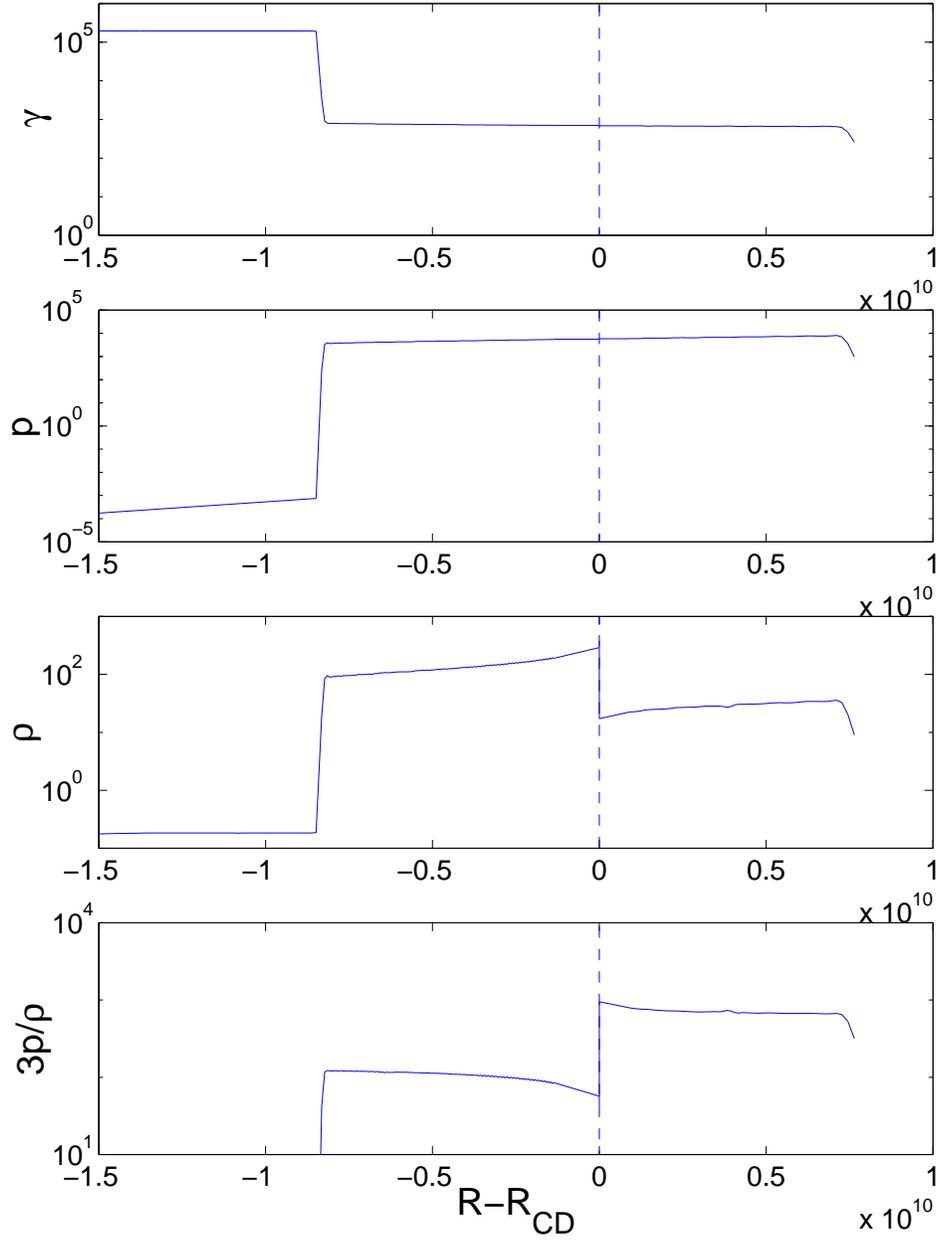,width=5in}} \vspace{10pt}
\caption{ Profile: Lorentz factor, pressure, density and
random Lorentz factor at the crossing time,
$R_{\Delta,num}$. 
The x-axis is the distance from the contact discontinuity.}
\label{fig:profile}
\end{figure}

 We numerically define the reverse shock crossing time $R_{\Delta,num}$
as the time at which the energy in the shocked shell becomes equal to 
the unshocked one. $R_{\Delta,num}$ is $\sim 3.3 \times 10^{16} \sim
0.7R_\Delta$ cm in this case. The profile at $R_{\Delta,num}$ is plotted
in figure \ref{fig:profile}, we can clearly see the forward and 
the reverse shock and the contact discontinuity. The widths of the
shocked ISM and the shocked fireball are comparable, the analytical
estimate is $\sim R_0/2=1.5 \times 10^{10}$cm.  

The Lorentz factor of the shocked shell is about 700 which is comparable
to the analytic estimate $\xi^{3/4}\eta \sim 1100$. There is a gap of
the density at the contact discontinuity while pressure is almost
constant through the shocked regions, then the random Lorentz
factors (or the temperature) $e/\rho$ are different in the two shocked
regions separated by the discontinuity. The numerical values are about 100
in the reverse shocked and 700 in the forward shocked while the analytic
estimates is $\xi^{-3/4}\sim 180$ and $\xi^{3/4}\eta \sim 1100$, respectively.

\begin{figure}[b!]
\centerline{\epsfig{file=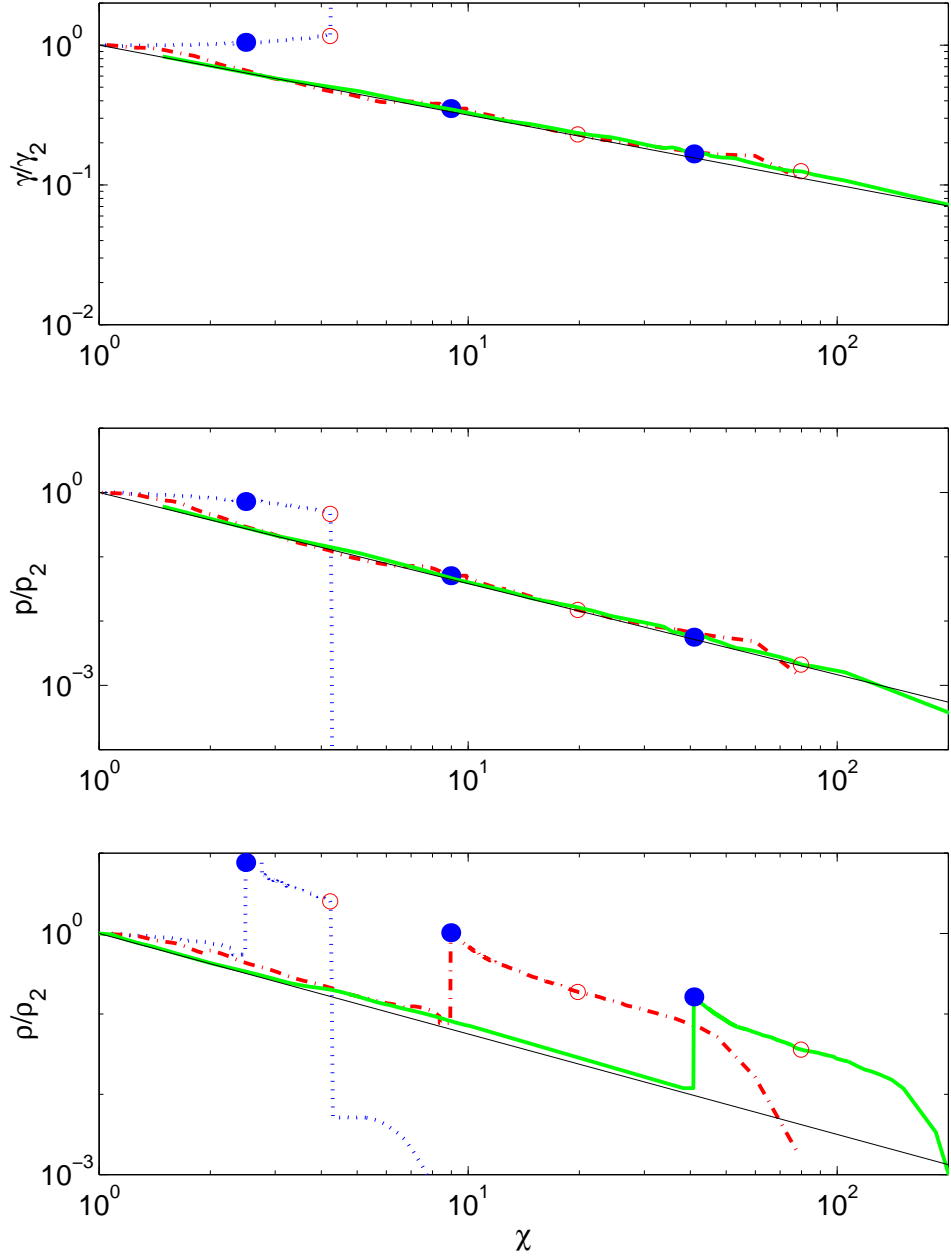,width=5in}} \vspace{10pt}
\caption{ RRS case: Lorentz factor, pressure and
density as functions of $\chi$ at $R_{\Delta,num}$ (dotted lines),
 at $2R_{\Delta,num}$ (dotted dashed lines) and at $3R_{\Delta,num}$
 (solid lines). The Blandford-McKee solution is the thin solid lines.
The ejecta shell is bounded by the contact discontinuity (filled circle) 
and the fluid element where the reverse shock is at $R_{\Delta,num}$ 
(open circle).
$\gamma$, $p$ and $\rho$ are normalized with $\gamma_2$, $p_2$ and 
$\rho_2$ which are numerical values just behind the forward shock.}
\label{fig:chi_rrs}
\end{figure}

In the BM solution, the Lorentz factor, the density and the pressure are 
given by  
\begin{equation}
\gamma/\gamma_2=\chi^{-1/2}, p/p_2=\chi^{-17/12}, 
\rho/\rho_2=\chi^{-5/4},
\end{equation}
where $\gamma_2$, $p_2$ and $\rho_2$ are the values of the Lorentz
factor, the pressure and the density just behind the forward shock and
the similarity variable $\chi$ is defined by $\chi \equiv
1+16\gamma_2^2(t)(1-r/R(t))$ with the shock radius $R(t)$. Radius $r$ 
and time in the observer's rest frame $t$ are independent coordinates
here.  After the reverse shock crossed the shell at $\sim
R_{\Delta,num}$, a rarefaction wave begins to propagate from the inner
edge of the shell, crosses the contact discontinuity and continues
towards the forward shock. Around $2 R_{\Delta,num}$, it reaches the 
forward shock, the separation between the contact discontinuity and 
the forward shock becomes comparable with the thickness of the blast 
wave $R/4\gamma_2^2$. Then, the profile of the forward shocked region
approaches to the BM solution. The profile normalized by the value
just behind the forward shock at different times are plotted in 
figure \ref{fig:chi_rrs}. The boundaries of the ejecta shell are 
indicated by the filled and the open circles. The former is the contact
discontinuity, the latter marks the fluid element in which the reverse
shock has been at time $R_{\Delta,num}$. 

\begin{figure}[b!]
\centerline{\epsfig{file=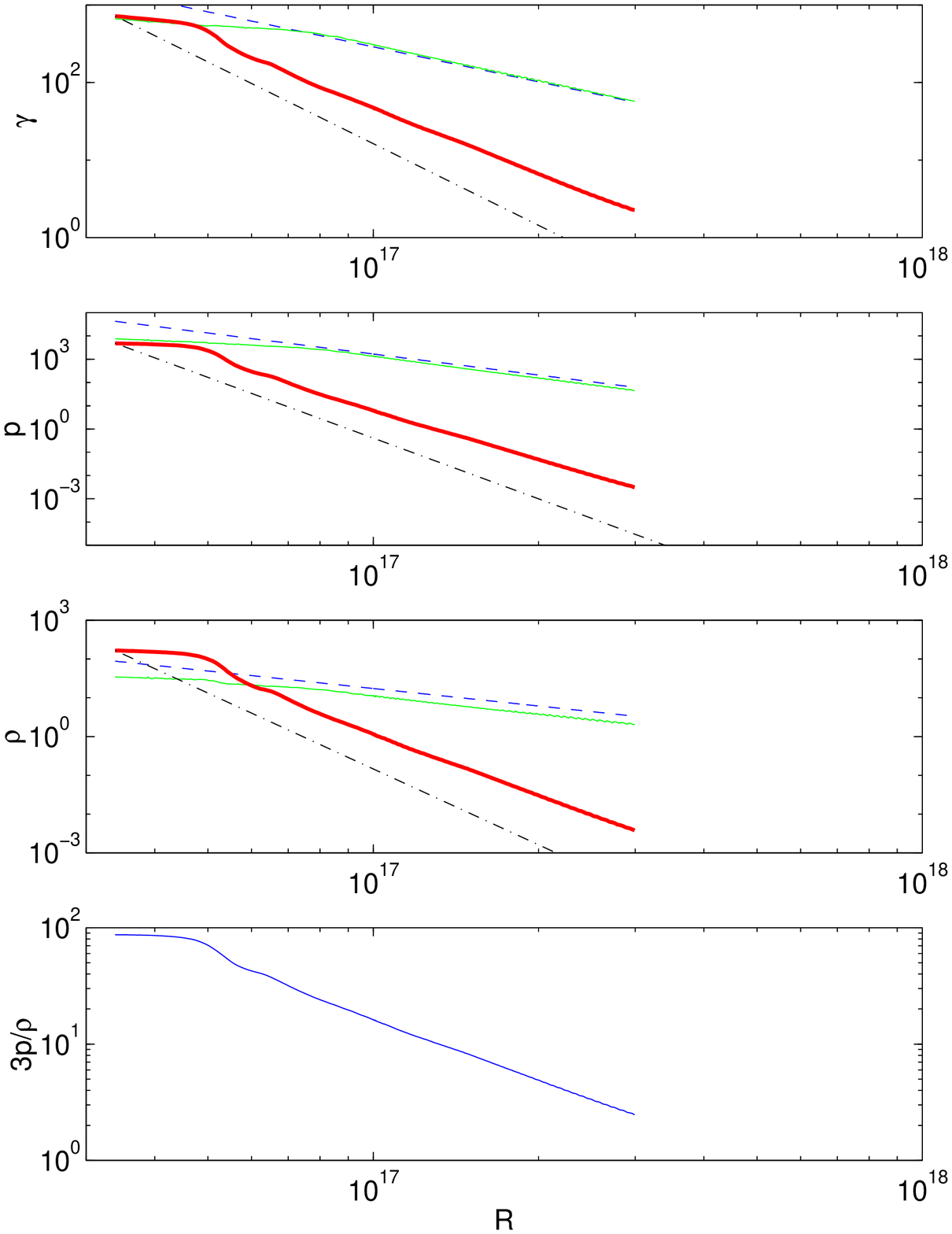,width=5in}} \vspace{10pt}
\caption{ 
Evolution of Lorentz factor, pressure and density: 
RRS case.
The numerical values just behind the forward shock are the thin solid lines,
the expected values from the Blandford-McKee solution are the thin
dashed lines. The numerical evolution of the averaged values in the ejecta 
shell (thick solid lines) are compared with the Blandford-McKee
scalings of a fluid element (thin dashed dotted lines).}
\label{fig:fs_rrs}
\end{figure}

The profile of the reverse shocked ejecta except the density one
(the right side of the filled circle) also approaches to the BM.
This can be understand as follows, since the temperature of the
shocked ejecta is relativistic, the internal energy dominates the
fluid inertia. The density gap at the contact discontinuity is
negligible for the evolution of the ejecta since its inertia is given
by its internal energy rather than its rest mass and the BM solution fits
the profile. The evolution of the Lorentz factor, the pressure and 
the density of the ejecta shell are plotted in Figure
\ref{fig:fs_rrs}. After the rarefaction wave reaches the forward shock
around $2R_{\Delta,num}$, the scalings of the ejecta are also adequately
approximated by the BM scalings for a fluid element.

\begin{figure}[b!]
\centerline{\epsfig{file=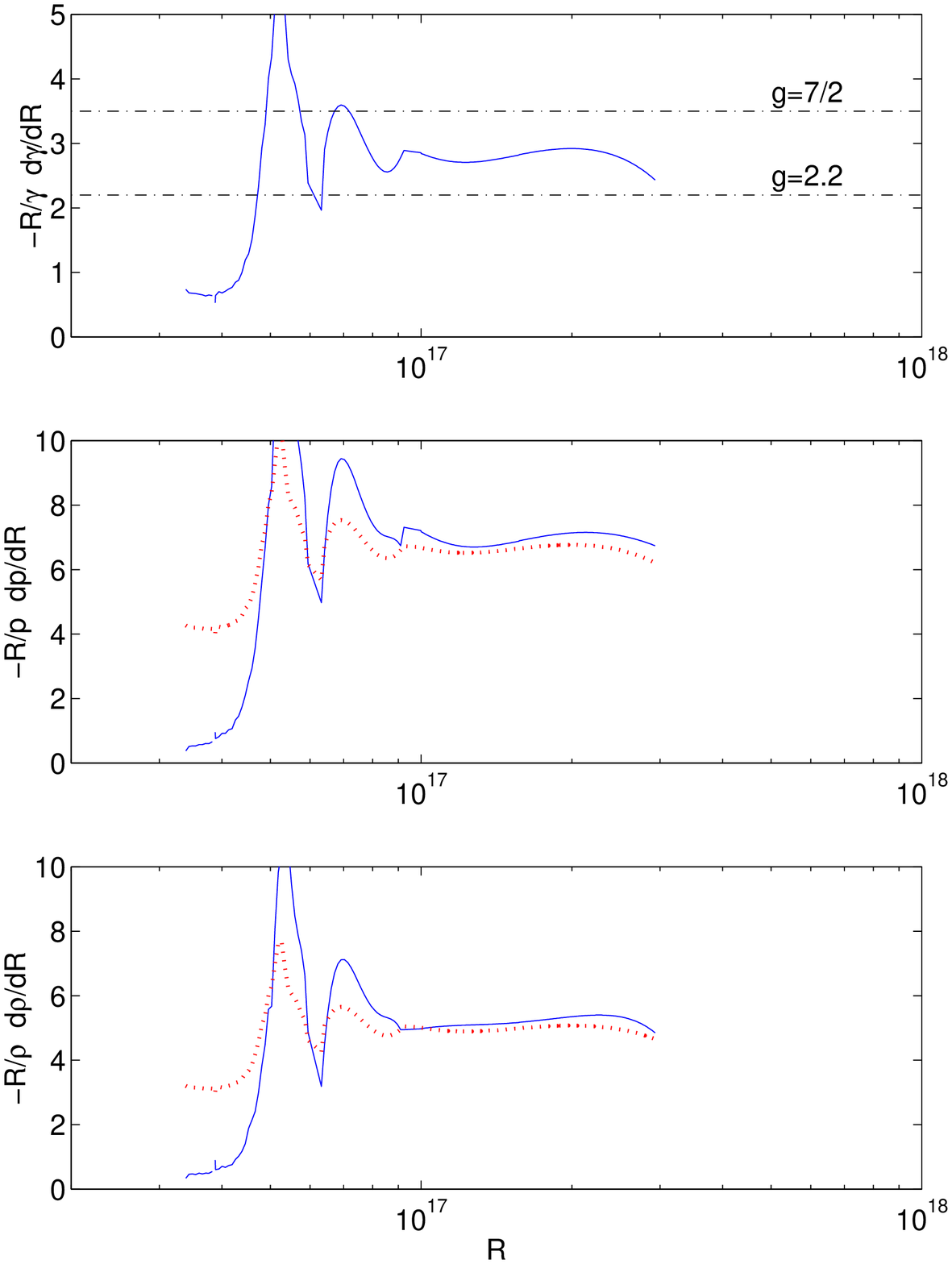,width=5in}} \vspace{10pt}
\caption{ Effective scaling indexes vs Radius of the shell:
RRS case.
The numerical results are the solid lines. The dotted lines
in the middle and the bottom panel are analytic estimates assuming
the numerical index $g$.}
\label{fig:slope_rrs}
\end{figure}

We define the effective scaling indexes, $-R/\gamma \cdot
d\gamma/dR$, $-R/p \cdot dp/dR$ and $-R/\rho \cdot d\rho/dR$.  The
evolutions of the averaged values in the shocked shell are plotted
against the radius of the forward shock in figure \ref{fig:slope_rrs}. 
The scaling index of the Lorentz factor $g$ is about $3$ in stead of $7/2$
of the BM solution, after the rarefaction wave reaches the forward shock.
The dotted lines depict the analytic estimate
(\ref{subrela}) assuming the numerical index $g$. The analytic
estimate fits the numerical results as long as the bulk motion is
relativistic, the spreading assumption and the adiabatic expansion
law are therefore good approximation.

\subsection{the Newtonian Reverse Shock Case}

We consider the other extreme case in which the shocked ejecta is cold
from the beginning. The initial condition is $E=3\times10^{54}$erg,
$\rho_1=10$ proton $\mbox{cm}^{-3}$, $\eta=10^3$ and $R_0=3\times
10^7$cm. This corresponds to the NRS case of $\xi=44$.

In the NRS case, the energy of the ejecta is transferred to the ISM
when the forward shock collects ISM mass of $E/\eta^2$ around $R_\gamma
\sim 6\times 10^{16}$ cm. The reverse shock crosses the ejecta at the
same time. The numerical estimate is $R_{\Delta,num} \sim 3 \times
10^{16}$cm. 

\begin{figure}[b!]
\centerline{\epsfig{file=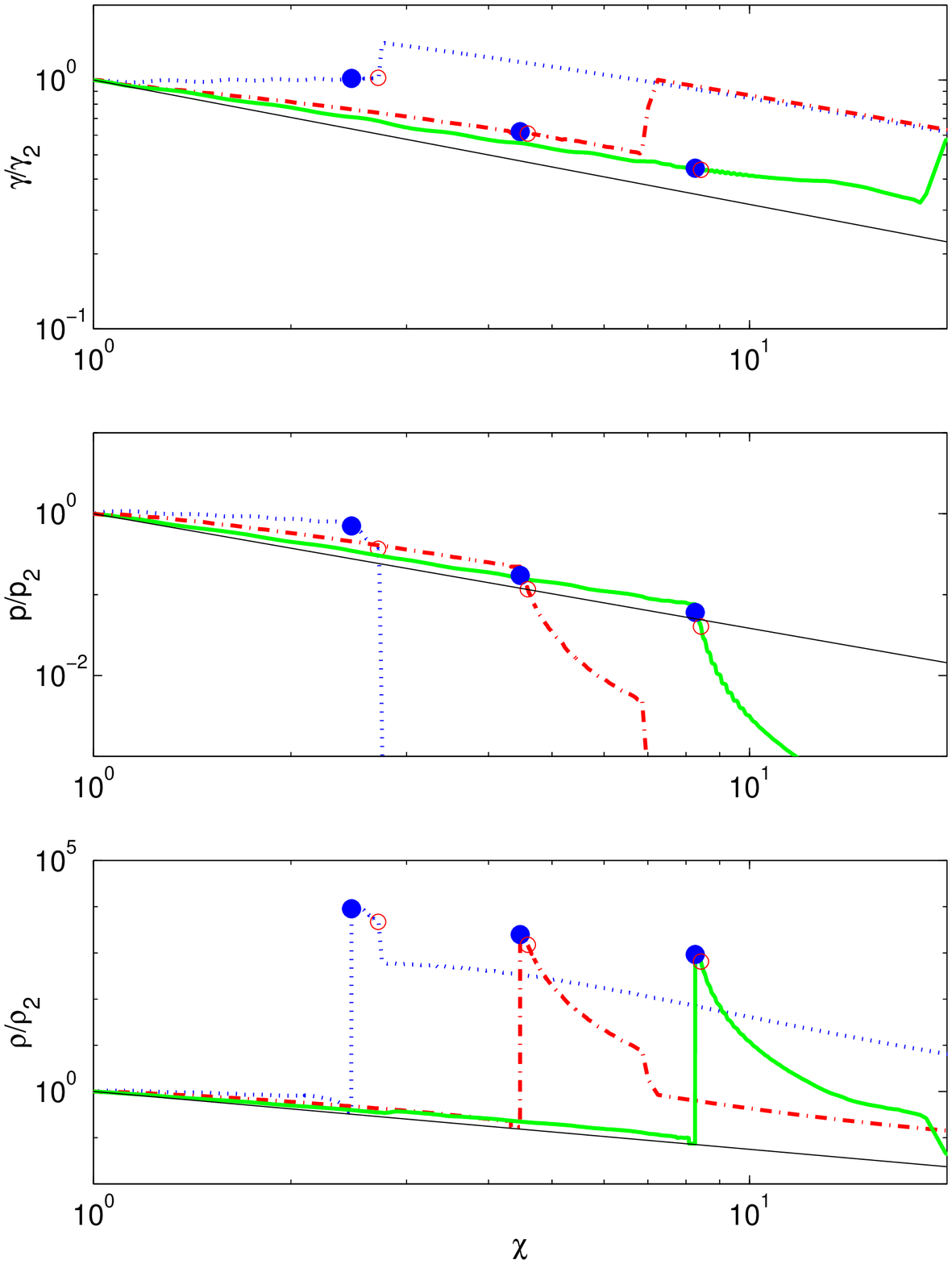,width=5in}} \vspace{10pt}
\caption{NRS case:
Lorentz factor, pressure and density as functions of
$\chi$ at $R_{\Delta,num}$ (dotted lines), at $2R_{\Delta,num}$ (dotted
 dashed lines) and at $3R_{\Delta,num}$ (solid lines). The
 Blandford-McKee solution is the thin solid lines.}
\label{fig:chi_nrs}
\end{figure}

The Lorentz factors of the shocked regions are about 700 at that time. 
The reverse shocked ejecta is already cold $3p/\rho \sim 0.05$ while the
forward shocked ISM is hot $3p/\rho \sim 700$. The profiles at different
times are plotted against the similarity variable $\chi$ in figure
\ref{fig:chi_nrs}. At $\sim 2R_{\Delta,num}$ the forward shocked region
is relatively well described by the BM scaling, but there is a break at
the contact discontinuity and the reverse shocked region deviates from
it. The reverse shock does not heat the shell well, the pressure of
the shocked region is less than the value expected from the BM
solution. The evolution of $\chi_e$ for a fixed element in the BM
solution is given by $\chi_e=\chi_{e0}(R/R_0)^4$ where $R_0$ and
$\chi_{e0}$ are the initial shock radius and the initial value of
$\chi$. At $R_{\Delta,num}$, the boundaries of the shell are at 
$\chi=2.5$ and $2.7$, the analytic formula gives that the values are 40 
and 43 at $2R_{\Delta,num}$, 203 and 219 at $3R_{\Delta,num}$. 
However, the shocked ejecta departs from the forward shock slower,
the numerical correspondences are 4.5 and 4.6 at $2R_{\Delta,num}$,
8.3 and 8.4 at $3R_{\Delta,num}$. 

\begin{figure}[b!]
\centerline{\epsfig{file=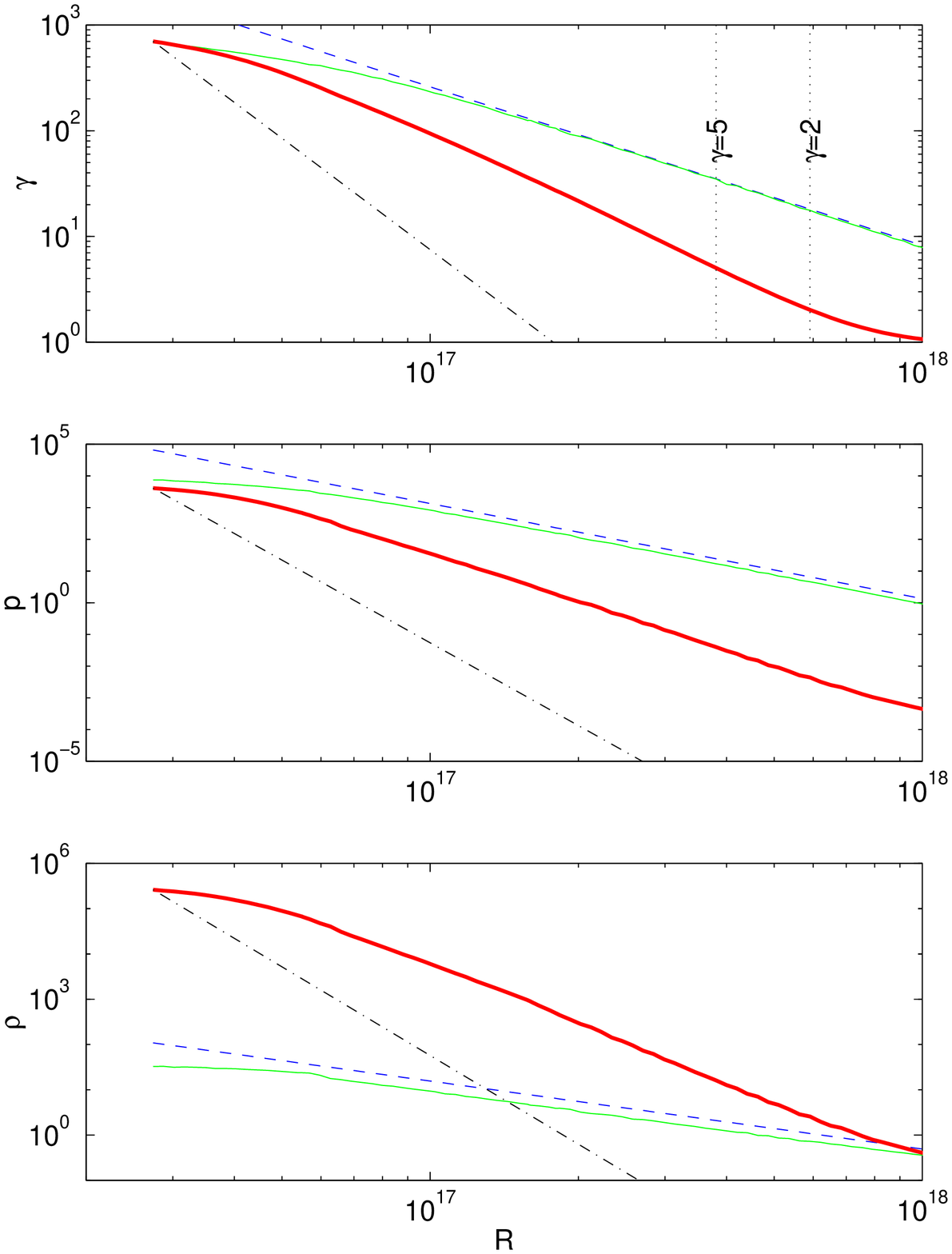,width=5in}} \vspace{10pt}
\caption{ 
Evolution of Lorentz factor, pressure and density:
NRS case.
The numerical values just behind the forward shock are the thin solid lines,
the expected values by the Blandford-McKee solution are the thin
dashed lines. The numerical evolution of the averaged values in the ejecta 
shell (thick solid lines) are compared with the Blandford-McKee
scalings of a fluid element (thin dashed dotted lines). The vertical
dotted lines show the radius where the Lorentz factor of the ejecta is 
2 or 5.}
\label{fig:fs_nrs}
\end{figure}

\begin{figure}[b!]
\centerline{\epsfig{file=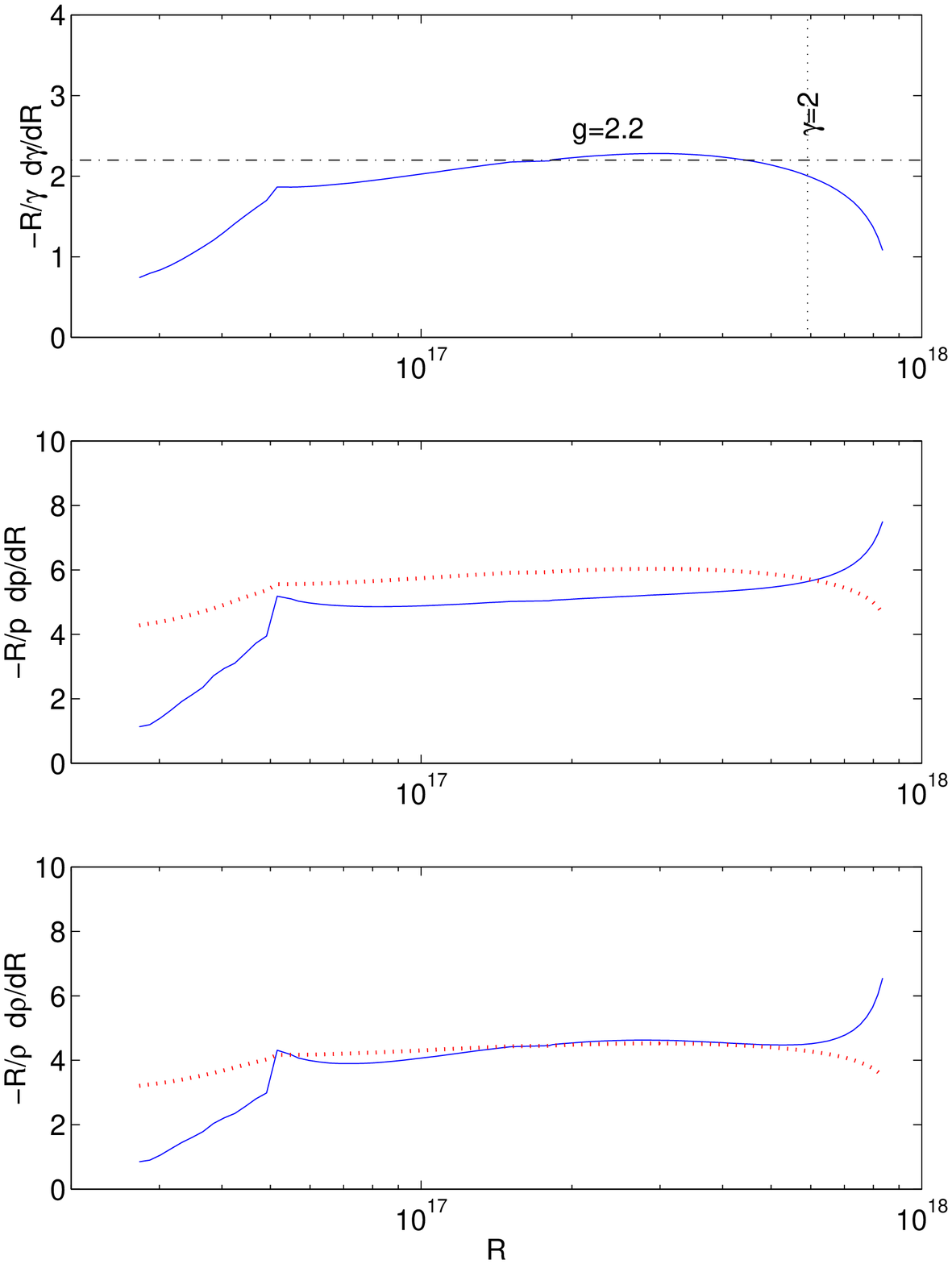,width=5in}} \vspace{10pt}
\caption{ Effective scaling indexes vs Radius of the shell:
NRS case.
The numerical results are the solid lines. The dotted lines
in the middle and the bottom panel are analytic estimates assuming
the numerical index $g$.}
\label{fig:slope_nrs}
\end{figure}

The profile of the shocked regions does not deviate from the self
similar scalings so much in figure \ref{fig:chi_nrs}, but the
evolution of $\chi_e$ is much slower. The hydrodynamic variables
of the shocked shell evolve slower. In figure \ref{fig:fs_nrs}
the averaged values of the Lorentz factor, the pressure and 
the density in the shell are compared with the BM solution (equations
\ref{bm_scalings}).  
Though the values just behind the forward shock (thin solid line) fit 
the BM scalings well around $2R_{\Delta,num}$,
the numerical scaling (thick solid line) is very different from the BM
one for a fluid element. The effective scaling index of 
$\gamma$ is about $2.2$ as we can see in fig \ref{fig:slope_nrs}.
The analytic estimates (\ref{subrela})
based on the spreading assumption and the adiabatic expansion 
are good approximation as long as the bulk motion is relativistic.

\subsection{GRB~990123}

We have seen the two extreme cases: the RRS ($\xi << 1$) and the NRS
($\xi >> 1$). In the RRS case the BM solution can describe well the
evolution of the shocked ejecta, but the scalings of the NRS are very
different from the BM. The fireball of GRB~990123 is actually a
marginal case as we see soon. However, such marginal cases behave
very much like the NRS as the shell becomes cold at the early stage.

According to the internal shocks model the duration of the gamma-ray
burst is determined by the thickness of the relativistic flow which is
the initial fireball size. The $T_{90}$ duration of GRB~990123
is 63 sec in the 50-300keV range (Galama et al. 1999). 
The burst profile is dominated by two 
peaks, each lasting 8 sec, separated by 12 sec, we assume 
$R_0/c = 40$ sec. The
observations suggest that the typical synchrotron frequency of the
reverse shock is below the optical bands quite early on, then 
the initial Lorentz factor of the ejecta is a few hundred (Sari 
and Piran 1999c). $\eta=400$ is assumed here.
The explosion energy and the ambient density are rather ambiguous.
We assume $E=1\times10^{54}$ergs and $\rho_1=5$ protons 
$\mbox{cm}^{-3}$. This is a marginal case of $\xi=0.7$.

The explosion energy might be higher since the internal shocks
can not convert the whole kinetic energy of the flow into the
thermal energy and that only some part of thermal energy goes into
random motions of the electrons (Kobayashi, Piran and Sari 1997). 
However, the sedov length determining
the hydrodynamic time scale are relatively insensitive to $E$. 
A non-spherical (jet) geometry of the ejecta are suspected from the
break in the afterglow light curve (Kulkarni et al., 1999). Even in this
case, a spherical fireball simulation is still valid to study the energy
transfer stage due to relativistic beaming.

When the reverse shock crosses the ejecta around
$R_{\Delta,num}\sim 8.4 \times 10^{16} cm \sim 0.9R_\Delta$,
The Lorentz factor of the shocked regions is about 120.
The reverse shocked regions is already cold from the 
beginning $e/\rho \sim 0.6$ while the forward shocked region is 
hot $e/\rho \sim 120$. The temperature ratio is $\sim 200$
close to the analytic estimate $\eta\xi^{3/4}\sim 300$.

\begin{figure}[b!]
\centerline{\epsfig{file=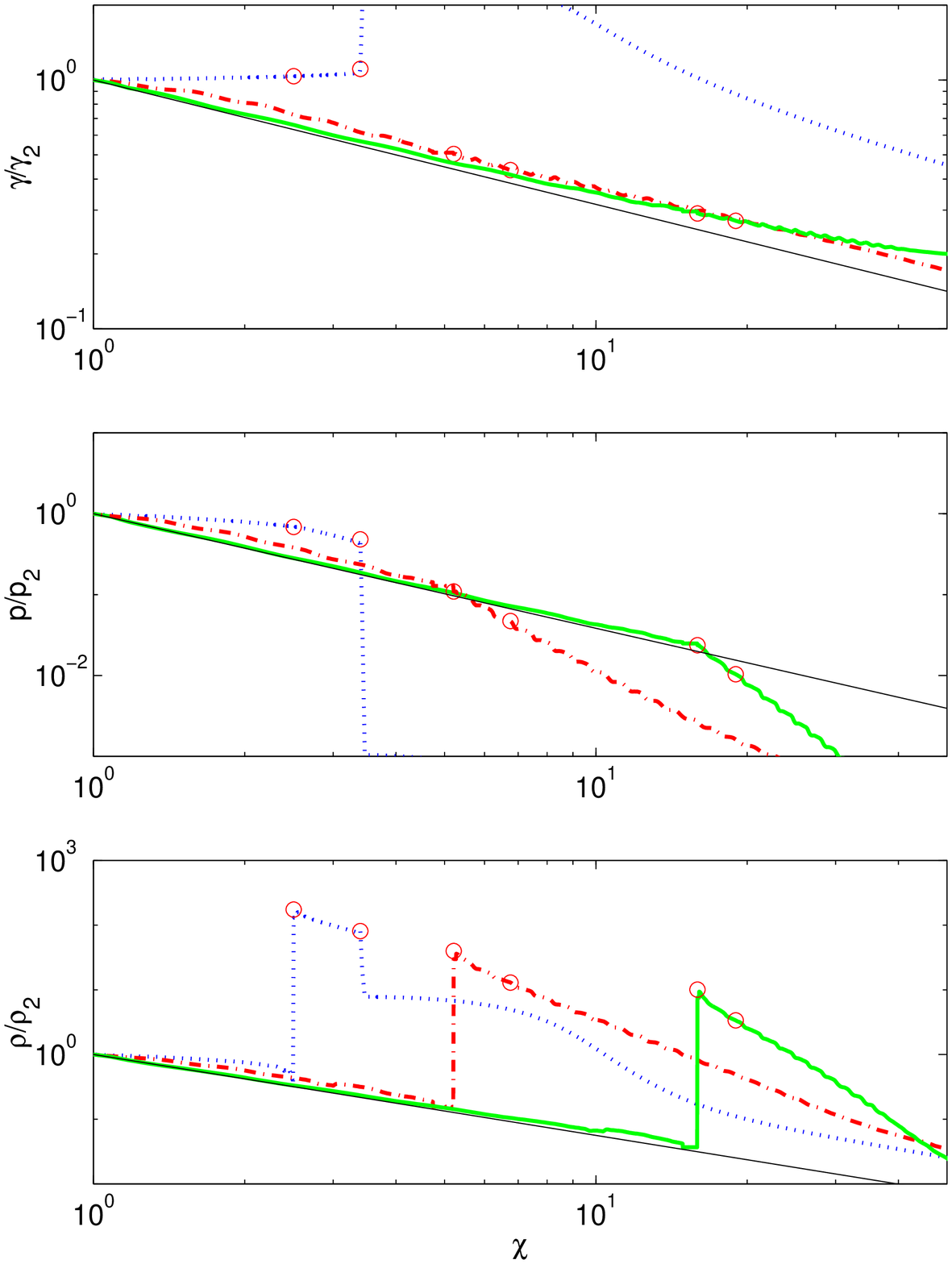,width=5in}} \vspace{10pt}
\caption{GRB990123:
Lorentz factor, pressure and density as functions of
$\chi$ at $R_{\Delta,num}$ (dotted lines), at $2R_{\Delta,num}$ (dotted dashed
lines) and at $3R_{\Delta,num}$ (solid lines). The Blandford-McKee solution 
is the thin solid lines.
}
\label{fig:chi_0123}
\end{figure}

At $\sim 2R_{\Delta,num}$ the forward shocked region is well described
by the BM scaling (see fig \ref{fig:chi_0123}), the reverse shocked 
region deviates from it. The reverse shock does not heat the shell 
as well as in the RRS case. At $R_{\Delta,num}$, the boundaries of the
shell are at $\chi=2.5$ and $3.4$, the analytic formula gives that the
values are 40 and 54 at $2R_{\Delta,num}$, 203 and 275 at
$3R_{\Delta,num}$. The numerical correspondences are 5.2 and 6.8 at
$2R_{\Delta,num}$, 16 and 19 at $3R_{\Delta,num}$, considerably smaller.

\begin{figure}[b!]
\centerline{\epsfig{file=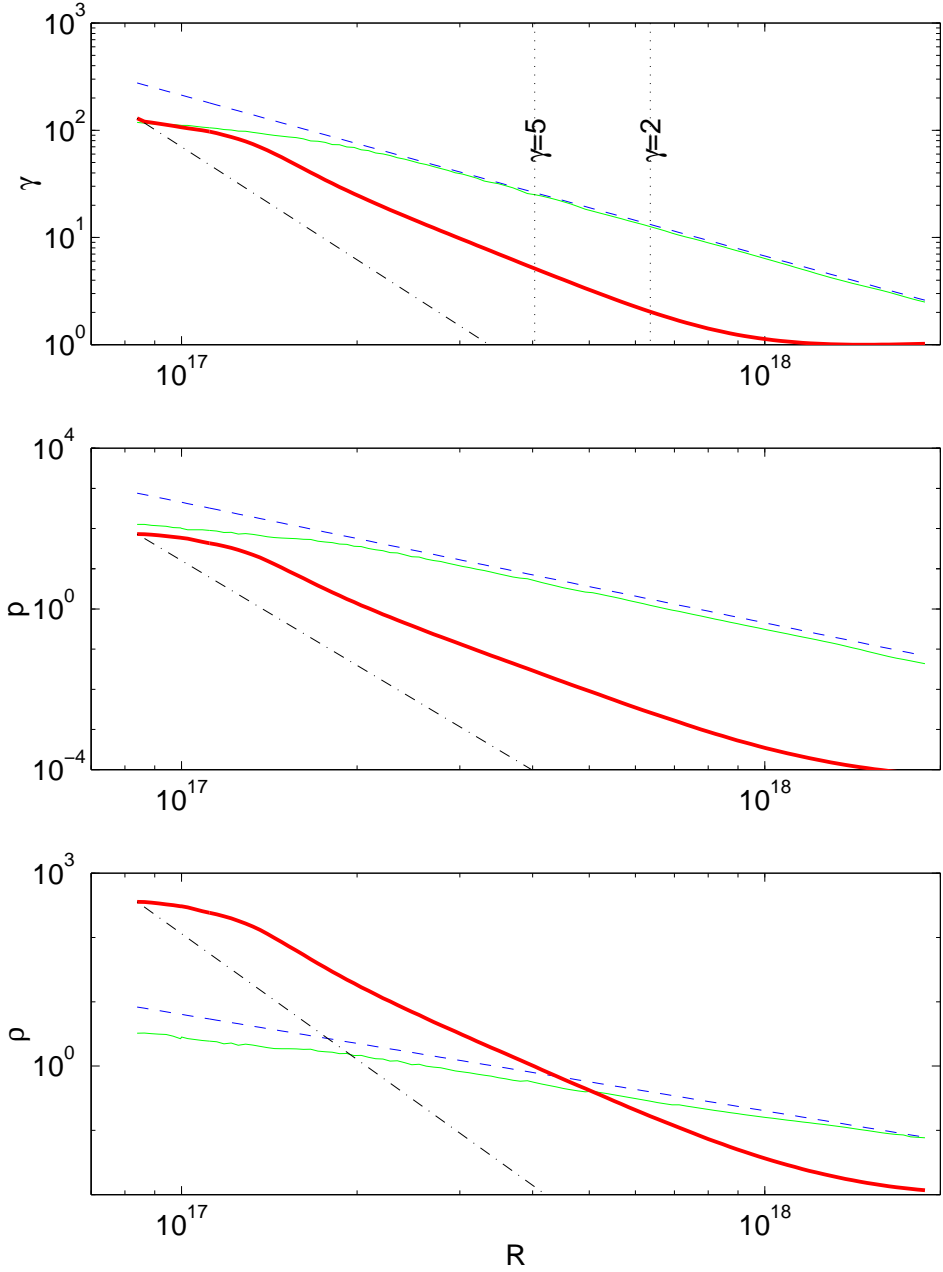,width=5in}} \vspace{10pt}
\caption{ 
Evolution of Lorentz factor, pressure and density:
GRB990123.
Numerical values just behind the forward shock (thin 
solid), expected values by the Blandford-McKee solution (thin
dashed). averaged values in the ejecta (thick solid lines) 
and the Blandford-McKee scalings of a fluid element (thin dashed 
dotted). The vertical
dotted lines show the radius where the Lorentz factor of the ejecta is 
2 or 5.}
\label{fig:fs_0123}
\end{figure}

\begin{figure}[b!]
\centerline{\epsfig{file=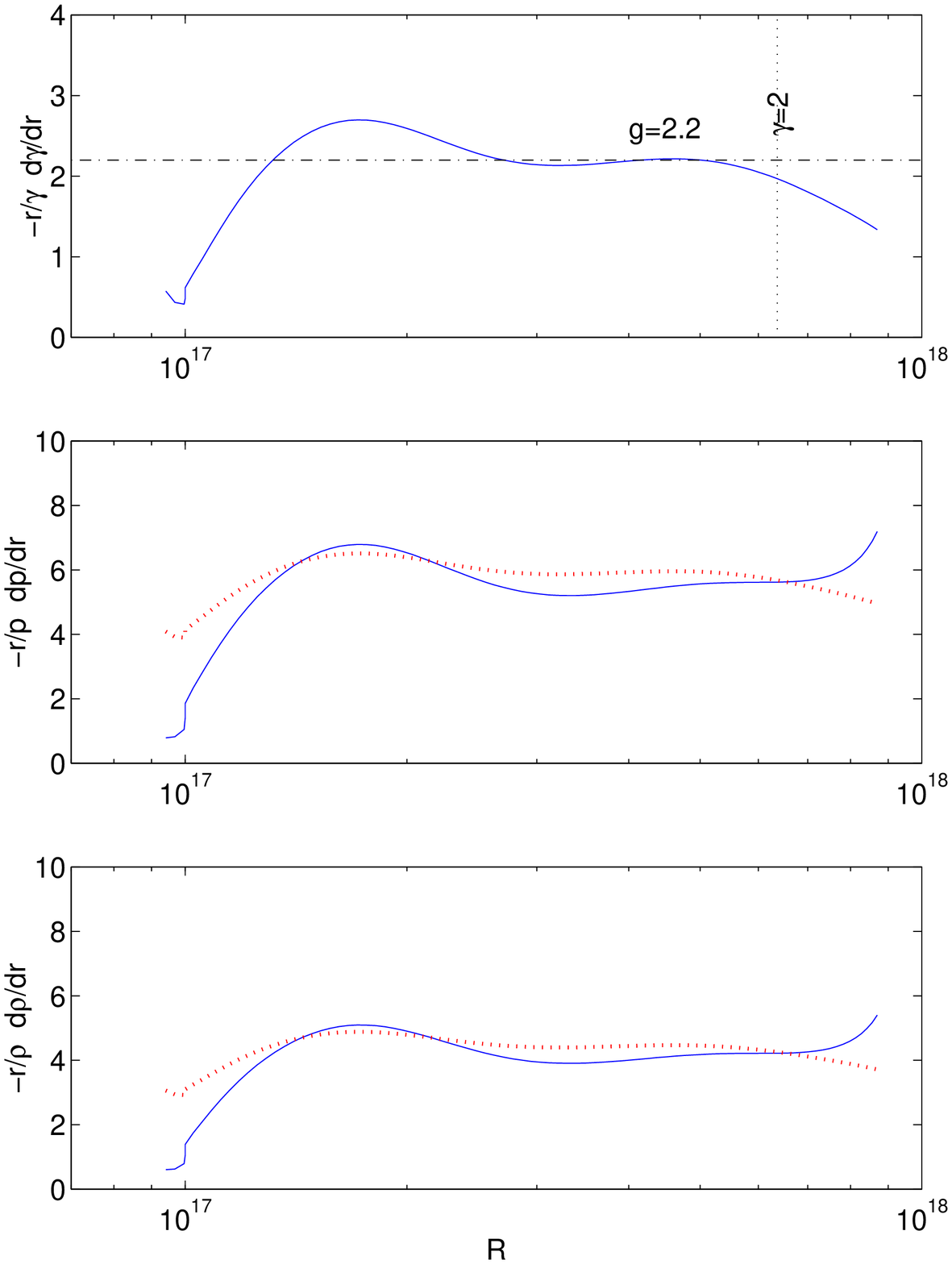,width=5in}} \vspace{10pt}
\caption{ Effective scaling indexes vs Radius of the shell:
GRB990123.
The numerical results are the solid lines. The dotted lines
in the middle and the bottom panel are analytic estimates assuming
the numerical index $g$.}
\label{fig:slope_0123}
\end{figure}

The hydrodynamic variables
of the shocked shell evolve slower than the BM (see \ref{fig:fs_0123}).
The vertical dotted 
lines show the time when the Lorentz factor of the ejecta shell
becomes $\gamma=5$ and $\gamma=2$. These happen at 0.55 day and 6.2
day respectively for the observer. The effective scaling index of 
$\gamma$ is about $2.2$ which is the same to that of the NRS case
(see fig \ref{fig:slope_0123}). The analytic estimates (\ref{subrela})
are good approximation again as long as the bulk motion is relativistic.

\begin{figure}[b!]
\centerline{\epsfig{file=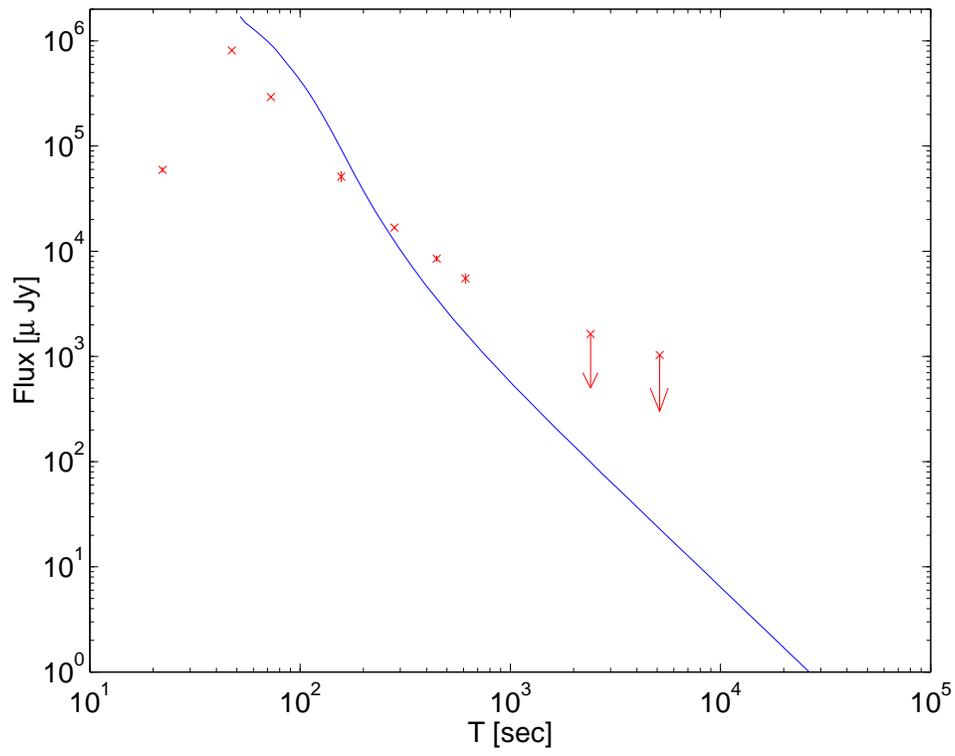,width=5in}} \vspace{10pt}
\caption{ Optical light curve: GRB990123.
The solid line depicts the numerical light curve.
The stars are the ROTSE observations and the arrows show the upper
 limits.}

\label{fig:lc}
\end{figure}

The optical flash of GRB~990123 initially decayed as $T^{-2}$.
This implies that the typical synchrotron frequency is
already below the optical band at the peak, we use 
the high frequency part ($\nu > \nu_m $) of the equation \ref{flux}
to estimate the optical flash.
Using a normalization $1.7$ Jy at the peak and the numerical
evolutions of $\gamma$, $p$ and $\rho$, the optical light curve 
is plotted in figure \ref{fig:lc}. The ROTSE  observations are 
also plotted (stars), the arrows mean the upper limits.
This normalization value of $1.7$ Jy is double of the second
ROTSE observation $0.81$ Jy, it gives good fits to
both of the optical flash and the radio flare. The normalization
is actually done at the radio flare.

The steepness of the light curve is
about $-2$ at late time with a steeper slope at early times. The 
numerical light curve reasonably fits the observation (see figure 
\ref{fig:lc}), the numerical light cure also qualitatively shows 
a steeper initial decline where the rarefaction wave is going 
through the shell.
The deviation can be due to some reasons. The second observation of
ROTSE has been assumed to be just at the peak, but 
the observation is sparse and the exposure duration is 5 sec. It is 
also possible that the real peak is in between the second (47 sec) 
and the third (72 sec) ROTSE data. This will allow the higher 
peak value. A more fundamental problem is that in internal shocks
model, the source is not a single explosion. At the end of the 
internal shock phase (GRB phase)
we are left with rather ordered flow in which faster ejecta are
the outermost ones and slower follow behind them. At the afterglow 
phase the outermost ones are decelerated by the ISM, the slower
ones collide into them. This effect makes the decay of the light curve 
slower (Panaitescu, M\'{e}szaros and Rees 1998).

Emission from the reverse shock can also explain the radio flare,
the radio detection one day after the burst (Sari and Piran
1999c, Kulkarni et al. 1999). The shocked ejecta initially
radiates in the optical band. As the ejecta expands, the 
temperature of the ejecta becomes lower. The emission frequency and 
the flux drop quickly, eventually the emission comes to the radio 
band and further lower. 

\begin{figure}[b!]
\centerline{\epsfig{file=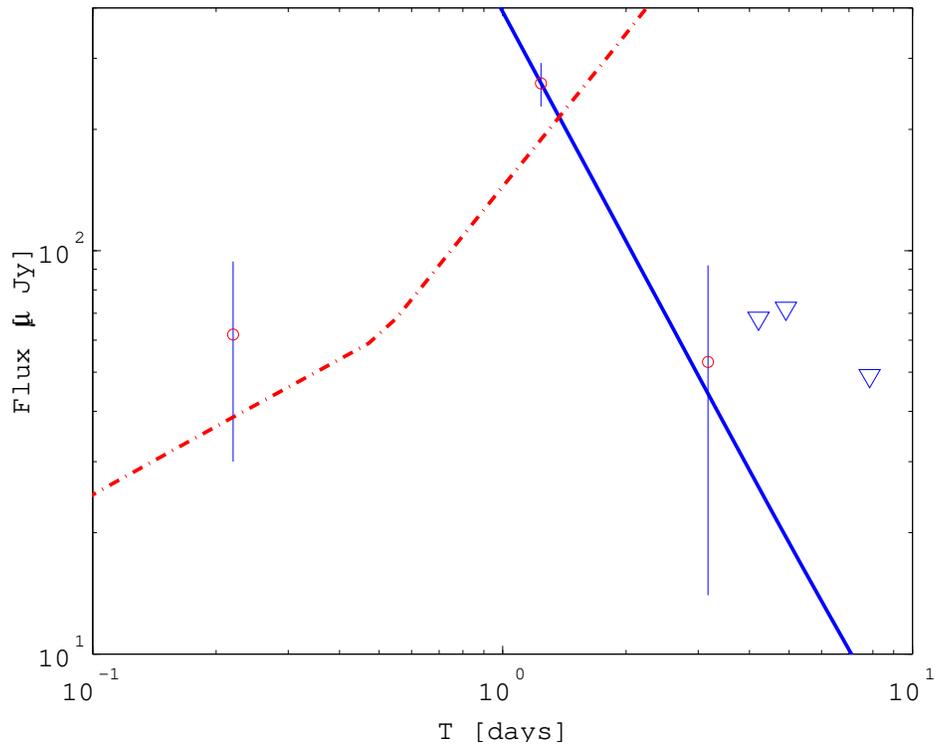,width=5in}} \vspace{10pt}
\caption{ Radio light curve: GRB~990123.
Numerical light curve (solid line), self absorption upper limit
(dotted dashed line). Observations are also plotted,
triangles are upper limits}
\label{fig:radio}
\end{figure}

We use the same normalization with the optical flash estimate, 
the flux of 1.7 Jy at the typical frequency $\nu_m=5\times 10^{14}$ Hz 
at 52 sec. The peak frequency $\nu_m \propto \gamma p^{5/2} \rho^{-2}$
reaches 8.5GHz after 12 hours, the peak flux $F_{\nu_m} \propto \gamma
p^{1/2}$ is  1.5 mJy at that time. After that the flux at the
radio band drops as 
$\propto \gamma ^{(\hat p+1)/2}p^{(5\hat p-3)/4}\rho^{-(\hat p-1)}$.
The flux at 8.5 GHz is plotted in figure \ref{fig:radio}. 

Self absorption can reduce the radio flux. Using equation 
\ref{bb} with $\epsilon_e = 0.6$ (Granot, Piran and Sari 1998), 
the upper limit is numerically 
estimated (dotted dashed line in fig \ref{fig:radio}). When 
accounting for this, the resulting emission is the minimum 
between the solid line and the dotted dashed line. The estimates 
fit very well to the radio observation. Since the emission is 
normalized with the observation, only the evolution scalings of the 
hydrodynamic variables are important to estimate it.
On the other hand, the self absorption estimate \ref{bb}
depends on the absolute values. It is relatively sensitive
to the Lorentz factor and the temperature of the ejecta.


\section{Discussion}


We have studied numerically the evolution of fireballs. Specifically
we focused on the evolution of the reverse shocked ejecta which is 
causing the optical flash and the radio flare
of GRB990123. We have seen that the Blandford-Mckee solution is 
not applicable to the shocked ejecta in a NRS case in which the 
reverse shock is Newtonian or mildly relativistic and the 
temperature of the shocked ejecta is not relativistic. However, 
in a RRS case in which the reverse shock is relativistic, the 
profile of the reverse shocked ejecta as well as the forward 
shocked ISM approached to the solution after the reverse shock 
had crossed the ejecta. The self similar solution is rigorous
only after the details of the initial conditions becomes no longer
important. We find that initially a steeper decline may be expected,
qualitatively compatible with the observations of GRB~990123.

The hydrodynamics of the cold shocked ejecta are very different from 
that of the hot shocked ejecta which is well described by the 
Blandford-Mckee solution. The numerical scaling relations for a cold
one were well approximated by the spreading assumption and the 
adiabatic expansion law, while
the Blandford-Mckee solution satisfies the relativistic version of it. 
We have estimated the radiation from the fireball ejecta 
in both cases: cold and hot shells. Surprisingly, we find that both
evolutions give rather similar light curves even though the
hydrodynamics is very different. 

After the reverse shock crossed the ejecta shell, the electrons cools
due to the adiabatic expansion, the emission frequency 
and the flux drop quickly. The radio flare, the single radio detection 
one day after the burst is the late time correspondence to the early
optical flash. Using the numerical results
we have shown that the reverse shock can explain the radio flare
observation a day after the burst, in agreement with the previous
analytical estimates.

The late time afterglow is less sensitive to the property of the
original ejecta, which is determined by only one parameters, the
ratio between the explosion energy and the density of the ambient
medium. The new observation window, the optical flash is very 
useful to probe the original ejecta and ultimately the inner engine
producing it. Future optical flash and radio flare observations will
enable us to know about the original ejecta more. 
Numerical simulations similar to the one presented here will be 
useful for the detailed study.

S.K. acknowledges support from the Japan Society for the Promotion 
of Science. R.S is supported by the Sherman Fairchild foundation.

\vspace{1cm} 
\noindent {\bf References}\newline
\noindent Akerlof, C.W. et al. 1999, GCN 205.\newline
Blandford,R.D. \& McKee,C.F. 1976, Phys. of Fluids, 19, 1130.\newline
Fenimore,E.E., Ramirez-Ruiz, E. and Wu, B. 1999, astro-ph/9902007.\newline
Galama, T.J., 1998, astro-ph/9804191.\newline
Galama, T.J., 1999, astro-ph/9903021.\newline
Granot,J., Piran,T. \& Sari,R. 1998, astro-ph/9808007.\newline
Kippen, R.M. et al., GCN, 224.\newline
Kobayashi,S., Piran,T. \& Sari,R. 1997, ApJ, 490, 92.\newline
Kobayashi,S., Piran,T. \& Sari,R. 1999, ApJ, 513, 669.\newline
Kulkarni, S.R., et al., Nature, submitted, astro-ph/9902272.\newline
M\'{e}szaros,P. \& Rees,M.J. 1997, ApJ, 476, 231.\newline   
M\'{e}szaros,P. \& Rees,M.J. 1999, MNRAS, submitted,astro-ph/9902367.\newline  
Panaitescu,A., M\'{e}szaros,P. \& Rees,M.J. 1998, ApJ, 503, 314.\newline  
Sari,R. \& Piran,T. 1995, ApJL, 455, 143.\newline
Sari,R. \& Piran,T. 1997, ApJ, 485, 270.\newline
Sari,R. \& Piran,T. 1999a, A\&A, in press, astro-ph/9901105.\newline
Sari,R. \& Piran,T. 1999b, ApJ, in press, astro-ph/9901338.\newline
Sari,R. \& Piran,T. 1999c, ApJL, in press,  astro-ph/9902009.\newline
\newpage 


\end{document}